# Uncertainty of Short-term Wind Power Forecasts - A Methodology for On-line Assessment.

George N. Kariniotakis, *Member, IEEE* and Pierre Pinson

*Abstract*— The paper introduces a new methodology for assessing on-line the prediction risk of short-term wind power forecasts. The first part of this methodology consists in computing confidence intervals with a confidence level defined by the end-user. The resampling approach is used for this purpose since it permits to avoid a restrictive hypothesis on the distribution of the errors. It has been however appropriately adapted for the wind power prediction problem taking into account the dependency of the errors on the level of predicted power through appropriately defined fuzzy sets. The second part of the proposed methodology introduces two indices, named as MRI and PRI, that quantify the meteorological risk by measuring the spread of multi-scenario Numerical Weather Predictions and wind power predictions respectively. The multi-scenario forecasts considered here are based on the 'poor man's' ensembles approach. The two indices are used either to fine-tune the confidence intervals or to give signals to the operator on the prediction risk, i.e. the probabilities for the occurrence of high prediction errors depending on the weather stability. A relation between these indices and the level of prediction error is shown. Evaluation results over a three-year period on the case of a wind farm in Denmark and over a one-year period on the case of several farms in Ireland are given. The proposed methodology has an operational nature and can be applied to all kinds of wind power forecasting models.

*Index Terms*— Wind power forecasting, confidence intervals, prediction risk, ensemble weather forecasts.

## I. INTRODUCTION

Nowadays, wind farm installations in Europe exceed 28 GW, while projections for year 2010 foresee an installed capacity up to 75 GW. Such a large-scale integration of wind generation causes several difficulties in the management of a power system. Predictions of wind power production up to 48 hours ahead are recognized as a major contribution to a secure and economic power system operation.

Apart from spot forecasts of wind power of major importance is to provide tools for assessing on-line the accuracy of these forecasts. Such tools are expected to be particularly useful in trading wind power in a liberalized electricity market since they can prevent or reduce penalties in situations of poor prediction accuracy. In practice today, uncertainty is given in the form of confidence intervals or error bands around the spot wind power predictions.

Typical confidence interval methods, developed for models like neural networks [1] are based on the assumption that prediction errors follow a Gaussian distribution. This however is not the case for wind power predictions, for which error distributions exhibit some skewness, while the confidence intervals are not symmetric around the spot prediction due to the shape of wind turbine power curves. Moreover, the level of predicted wind speed introduces some non-linearity to the estimation of the intervals; i.e. at the cut-off speed, the lower power interval may switch to zero due to the cut-off effect. The limits introduced by the wind farm power curve (min, max power) are taken into account by the method proposed in [2], which is based on modeling errors using a β-distribution, the parameters of which have to be estimated by a post-processing algorithm. This approach however is applicable only to "physical"-type [6] of models since such models estimate power using an explicit wind turbine power curve - i.e. a function that gives the wind power output from the wind speed at the turbine level, which is not necessarily the case for statistical or artificial intelligence based models [3].

In [4], wind speed errors are classified as a function of look-ahead time and then they are transformed to power prediction errors using the wind turbine power curve vs wind speed. This method, whose main drawback is the symmetry of the intervals, is also limited for application to physical models rather than statistical ones since it requires wind speed predictions at the level of the wind farm. Furthermore, it does not provide uncertainty as a function of a pre-specified confidence level. The wind speed errors are estimations provided by the Numerical Weather Prediction (NWP) model. As a consequence, this method does not take into account the modeling error itself that might be due to the spatial refinement of weather predictions or to the power curve used.

In a follow-up paper [5], the authors show a relation between specific meteorological patterns (defined from measurements) and various levels of forecasting error: this is a first step in the definition of risk indices in order to quantify the weather predictability.

A model performance evaluation is usually done on a "global" basis, i.e. over a long period of time. However, this performance is highly variable in time. The aim here is to develop a methodology for assessing the prediction accuracy in a more dynamic way. Such methodology is characterized by two different concepts: the *uncertainty* and *prediction risk* estimation. The first one corresponds to a visualization of the

Manuscript received February 23, 2004. This work was supported in part by the European Commission in the frame of the research project ANEMOS (ENK-CT-2002-00665).

G. N. Kariniotakis is with the Center for Energy Studies of Ecole des Mines de Paris, B.P. No 207, 06904 Sophia Antipolis France (phone: +33-493957501, fax: +33-493957535; e-mail: georges.kariniotakis@ ensmp.fr).
P. Pinson is with Center for Energy Studies of Ecole des Mines de Paris, B.P. No 207, 06904 Sophia Antipolis France (e-mail: pierre.pinson@ensmp.fr).

error distribution on an a-posteriori basis, while the second one "forecasts" the uncertainty and extreme errors as a function of the expected weather stability. The methodology includes the following three parts:

- <Uncertainty>: Development of confidence intervals for the spot power prediction. The approach is based on the resampling method, which is applied to samples of errors. Errors are classified using fuzzy sets to account for the level of power and the risk for cut-off events.
- <Prediction Risk>: Given that confidence intervals are estimated based on the past performance of the model, the second objective consists in developing additional preventive tools able to assess on-line the prediction risk as a function of the forecast weather situation. This is done through the development of on-line prediction risk indices based on ensembles of NWPs and wind power forecasts. These indices permit to derive rules for assessing the probability of high or extreme prediction errors due to unstable weather situations. These rules aim is to provide comprehensive information to the operators so that they are able to adjust the risk they are going to face when managing the predicted wind power, i.e. take low risk when forecast weather situation is unstable.
- Dynamic fine-tuning of the size of the intervals depending on the weather stability. This permits to avoid excessive risk or to take preventive actions in situations where high errors are expected.

The proposed methodology is applicable to both "physical" and "statistical" wind power forecasting models, while no hypothesis is made about the distribution of the prediction errors. It accounts for both modeling errors and errors based on the NWPs. It uses past wind power data, which are often available on-line by a Supervisory Control and Data Acquisition (SCADA) system, as well as NWPs, which are nowadays the basic input to all models.

The paper presents detailed results on case studies in Ireland and Denmark, where the aim is to predict the output of several wind farms for 48 hours ahead using on-line measurements and predictions from Hirlam NWP system. The evaluation is based on several years of data.

II. ESTIMATING THE UNCERTAINTY OF WIND POWER FORECASTS.

Let us define the prediction error for the look-ahead time $t+k$ as following:

$$e_{t+k/t} = p_{t+k} - \hat{p}_{t+k/t} \qquad (1)$$

where $\hat{p}_{t+k/t}$ is the forecast for look-ahead time $t+k$ produced by the model at time origin $t$, and $p_{t+k}$ is the measured wind power. The forecasted power is the *average power* the farm is expected to produce during the considered period if it would operate under an equivalent constant wind. As a consequence, intra-hourly variations of power and their impact are not considered. This convention comes from the fact that NWPs of wind speed are given as constant values for the step ahead considered (i.e. next hour). Following this convention, in practice the value for the measured power $p_{t+k}$ is derived from higher resolution measurements (i.e. each 1 min or 10 min etc.), which can be instantaneous values or energy ones depending on the acquisition system. The prediction error in (1) can vary between $\pm 100\%$ of the nominal wind park power $P_n$. For a non-bounded prediction model it can take values even outside this range. The observed prediction error itself is in general the result of three factors: a modeling error, an error due to the accuracy of the NWPs and finally, a stochastic component linked to the process itself.

*A. Pre-processing Based on Fuzzy Set Modeling.*

The first step before computing confidence intervals is to collect the prediction errors the model made in the past. The intervals that are going to be computed will rely on the most recent information on the model's performance. For this, a window in the past (a certain number of hours) is defined and used as a sliding window for storing the errors. Its size defines the size of the samples of errors. A separate sample is developed for each prediction horizon $k$ (i.e. for 1-hour ahead, 2-hour ahead, and so on), because the shape of the error distributions depend on the look-ahead time [7].

The power prediction errors depend on the errors involved in the prediction of wind speed by the NWP system [4], [8]. Due to its shape, the wind park power curve is able to amplify (between cut-in and rated speed) or to reduce (below cut-in speed or between rated and cut-off speed) the uncertainty introduced by the NWPs. Moreover, the non-linearity of the energy conversion process leads to a bias for certain ranges of forecast values: there is a trend to under-predict when predicting very low power output and a trend to over-predict when forecasting power close to the rated capacity. To account for these effects, the wind power curve is divided into three ranges of power: low, medium and high, which are characterized by fuzzy sets. The prediction errors are then classified in samples $S^1_{k,p}$, $S^2_{k,p}$ and $S^3_{k,p}$ depending on the predicted power range (Fig. 1). Hence, the confidence interval estimation is carried out using the error samples corresponding to the power class of the forecast power.

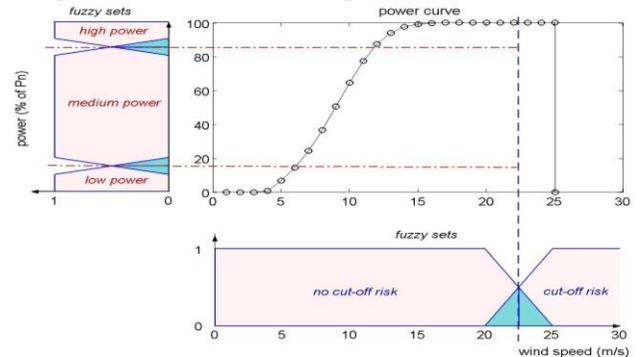

Fig. 1: *Splitting the power curve into three power class fuzzy sets and two cut-off risk zones.*

In a similar way, in order to deal with the risk due to the cut-off event, the range of wind speed values is divided into two ranges corresponding to a "no cut-off risk zone" for low wind speeds, and to a "cut-off risk zone" for wind speeds close or higher than cut-off. Like for the predicted power, errors are stored in samples $S_{k,ws}^1$ and $S_{k,ws}^2$, depending on the cut-off risk. An appropriate fuzzy set is associated to each zone as shown in Fig. 1.

*B. Confidence Interval Estimation by the Resampling Approach.*

Here is a formal definition of confidence intervals: the interval $I(X)$ computed from the sample data $X$ which, were the study repeated multiple times, would contain $(1-\alpha)\%$ of the time the true effect $x$, (1-$\alpha$) being the confidence level:

$$P(x \in I(X)) = P(x \in [z_{\alpha/2}, z_{1-\alpha/2}]) = 1-\alpha \quad (2)$$

A given set of observations (the sample) is a part of a whole population and can be seen as representative. The aim of methods like resampling is to have a better idea of the population distribution by going through the sample a high number of times. This evaluation of the population distribution can serve to estimate a mean, a variance, and even confidence bounds. No assumption is made concerning the distribution: that is the main reason why resampling is preferred to other methods like the ones based on simple standard deviation.

This procedure requires that both the sample size $N$ and the number of resampling loops $N_{loop}$ are significantly large. As a matter of fact, one always gets a new sample that is close to the original one, and the whole population distribution is not really simulated by this way. But, by calculating the mean of respectively the $\alpha/2\%$ lowest and $(1-\alpha/2)\%$ highest value of these randomly created samples, good estimates of the confidence limits $z_{\alpha/2}$ and $z_{1-\alpha/2}$ for the mean $m$ can be computed [9], [10].

In the case of wind power forecasting, the resampling method is applied by considering error samples defined as a function of the look-ahead time, the power range and the wind speed range. For a given horizon $k$, Fig. 1 represents the splitting of the predicted power range of values into three fuzzy sets $A_{k,p}^1$, $A_{k,p}^2$ and $A_{k,p}^3$, accounting respectively for low, medium and high predicted power. In a similar way, the forecast wind speed values allow to define two fuzzy sets $A_{k,ws}^1$ and $A_{k,ws}^2$, for situations without or with a risk of cut-off event.

In order to account for the specific shape of the power curve, the first step of the Resampling method is adapted by using fuzzy rules, in order to create a new sample that reflects the current conditions. A fuzzy rule will have the form:

$$\begin{aligned} &\text{IF } \hat{p}_{t+k/t} \in D(A_{k,p}^1) \text{ and } \hat{ws}_{t+k/t} \in D(A_{k,ws}^1) \\ &\text{THEN } X \in IR^N, \ X \subset (S_{k,p}^1 \cap S_{k,ws}^1) \end{aligned} \quad (3)$$

where $D(A)$ stands for the support of the fuzzy set $A$. This rule means that if the predicted power is in the low range and the forecast wind speed in the "no cut-off risk" range, then the new generated sample $X \in IR^N$ will be composed by values picked in the intersection of the error samples accounting for these specific situations: $X$ will be generated by selecting randomly and with replacement $N$ values out of the intersection of $S_{k,p}^1$ and $S_{k,ws}^1$.

Since there are three fuzzy sets defined for the predicted power and two for the forecast wind speed, six rules of that kind can be formulated. However, only $N$ values have to be picked out to create a new sample. Therefore, the fuzzy set membership functions are used to define the share of each rule in the final sample. This consideration leads to a new form of the fuzzy rule:

$$\begin{aligned} &\text{IF } \hat{p}_{t+k/t} \in D(A_{k,p}^i) \text{ and } \hat{ws}_{t+k/t} \in D(A_{k,ws}^j) \\ &\text{THEN } X^{ij} \in IR^{N^{ij}}, \ X^{ij} \subset (S_{k,p}^i \cap S_{k,ws}^j) \end{aligned} \quad (4)$$

with

$$N^{ij} = \frac{\mu_{k,p}^i(\hat{p}_{t+k/t}) \cdot \mu_{k,ws}^j(\hat{ws}_{t+k/t})}{\sum_{l=1}^{3}\sum_{m=1}^{2} \mu_{k,p}^l(\hat{p}_{t+k/t}) \cdot \mu_{k,ws}^m(\hat{ws}_{t+k/t})} N \quad (5)$$

and $i \in \{1,2,3\}$, $j \in \{1,2\}$. In this expression, $\mu_{k,p}^i(\cdot)$ and $\mu_{k,ws}^j(\cdot)$ are the membership functions of respectively the $i^{th}$ fuzzy set associated to power and the $j^{th}$ fuzzy set associated to wind speed.

Then, the generated sample will be composed by the sub-samples created by all the rules. However, we make the assumption that when the forecast wind speed is very high, showing a risk for a cut-off event, the value of the predicted power will not have a significant influence on that risk. Thus, the three rules corresponding to the cut-off risk are gathered to form only one rule. Finally the created sample consists in :

$$X = [X^{11}, X^{21}, X^{31}, X^{\cdot 2}]^T, \ X \in IR^N \quad (6)$$

where $X^{\cdot 2}$ denotes the sub-sample obtained with the unique cut-off risk rule defined before.

The adapted Resampling method proposed here is summarized by Algorithm 1. For every lead time, the confidence intervals are generated using this method, assuming that the prediction error the model makes is the mean of a distribution and that we would like to compute confidence intervals for that mean.

**Algorithm 1:** Calculate confidence bounds $z_{\alpha/2}$ and $z_{1-\alpha/2}$ by an adapted Resampling approach

$z_{\alpha/2} \Leftarrow 0$, $z_{1-\alpha/2} \Leftarrow 0$
   **for** $i = 1$ to $N_{loop}$ **do**
      *Creation* of a new sample $Y_i$ using the fuzzy-rule based method described by (5), (4) and (6).
      Sort $Y_i$ in ascending order
      $z_{\alpha/2}^i \Leftarrow$ the $N(\alpha/2)^{th}$ value of $Y_i$
      $z_{1-\alpha/2}^i \Leftarrow$ the $N(1-\alpha/2)^{th}$ value of $Y_i$
   **end for**
$z_{\alpha/2} \Leftarrow \frac{1}{N_{loop}} \sum_{i=1}^{N_{loop}} z_{\alpha/2}^i$,    $z_{1-\alpha/2} \Leftarrow \frac{1}{N_{loop}} \sum_{i=1}^{N_{loop}} z_{1-\alpha/2}^i$

III. PREDICTION RISK ASSESSMENT BASED ON ENSEMBLE FORECASTS.

Low quality forecasts are due partly to the power prediction model, and partly to the numerical weather prediction system. Indeed, an unstable atmospheric situation can lead to very poor numerical weather predictions and thus to worthless wind power ones. In contrast, when the atmospheric situation is stable, one can expect more accurate predictions for power.

In the following paragraphs the aim is to exploit the information included in the NWPs (and not in the measurements) in order to develop tools for on-line estimation of the meteorological risk in power predictions.

*A. Wind Speed Ensemble Forecasts for the Assessment of Weather Stability*

Meteorological Centres are able to produce different scenarios of Numerical Weather Predictions by perturbing the initial conditions of the forecasting model or by using different NWP models. These scenarios are called ensemble forecasts and permit to evaluate the stability of the weather regime as well as the meteorological predictability [11]. Both the U.S. National Center for Environmental Prediction (NCEP) and the European Center for Medium-Range Weather Forecasts (ECMWF) have produced operational ensembles for more than ten years.

The NWP wind speed prediction error is composed by a part that is independent of the lead-time and by an error that has a linear growth with the prediction horizon. The first includes the effects of weather disturbances that are smaller than the NWP resolution, while the second is due partly to the NWP model errors and partly to an error in the estimation of the initial state. Ensemble forecasts permit to assess the influence of this misestimating of the initial state in the weather forecasting evolution, and thus to quantify the prediction uncertainty [12].

For wind power applications only one set of NWP forecasts for the next 48 hours is often made available (or purchased) at a given time (i.e. Hirlam gives a unique 48-hour ahead forecast every 6 hours). Nevertheless, for a given hour, several predictions can be available from different time origins in the past (-6 hours, -12 hours, -18 hours, etc.). This kind of ensembles is known as "poor man's" ensemble forecasts. In a stable and well-predicted weather situation it is expected that these predictions will not differ significantly. Comparing all the available forecasts for the considered period one can assess weather stability and predictability.

Because we want to have a general evaluation of that stability, 4 sets of predictions of various ages (0, 6, 12 and 18 hours) for the following 24 hours are compared. Fig. 2 depicts the examples of a stable atmospheric situation (left picture, the forecasts are quite close) and of an unstable one (right picture, spread forecasts).

*B. Relation between Weather Stability and Wind Power Prediction Error*

There are several possibilities to measure the spread of the various weather forecasts. In [8] the standard deviation of the forecasts for each time-step is mentioned as an example. Our aim here is to evaluate the global atmospheric situation. This is why a unique representative index is defined for the following $N_h$ hours instead of indexes for every look-ahead time. In order to calculate the distance between two sets of forecasts, the authors have proposed in [7] a kind of Euclidian distance between the $N_h$-valued vectors containing the predicted wind speed for the $N_h$ following hours. Focus is given to the spread of wind speed forecasts because this variable is the most sensible input to wind power prediction models.

Then, an index, called hereafter "meteo-risk" MRI-index, is defined in [7] to measure the spread of the weather forecasts at a given time. This index uses the most recent forecast as a reference and reflects the variability of the older forecasts.

In the frame of the case studies of the paper, the horizon $N_h$ for the calculation of the MRI-index is set to 24 hours. Since Hirlam forecasts are provided every 6 hours, there are four sets of wind speed predictions covering the period. However, the same methodology could be applied to seven available sets of Hirlam forecasts on a 6-hour period for instance.

The MRI-index can be used to describe the distribution of weather situations as shown in [7]. Fig. 2 shows the patterns of a typical "stable" (MRI=0.3) and an "unstable" (MRI=2.9) situation.

The link of prediction error to weather stability is shown also for the second case study in Denmark. The power prediction errors, as obtained by the Fuzzy-Neural Network (F-NN) prediction model described in [3], are collected for a period covering 3 years. For the same period the MRI-index is estimated. By binning the data, calculating the average error $e_t^{24}$ (defined by (9)) for the next 24 hours for each bin, and comparing these averages to the global prediction error $\overline{e_t^{24}}$ of the model (defined by (10)) the representative points in Fig. 3 are obtained. This figure exhibits a roughly linear trend: the prediction error tends to increase linearly with the MRI-index: the tighter the Hirlam predictions are, the more accurate the wind power prediction model is. A linear fitting gives the solid curve shown in Fig. 3. We mention that the relation between the prediction error and the meteorological risk index is a trend because it is not possible to link directly a MRI-index value to an error value, though we can say that for low or high MRI values, there are respectively less and more chances for high prediction errors (see also Section IV). Making this assumption would mean that the prediction error the model

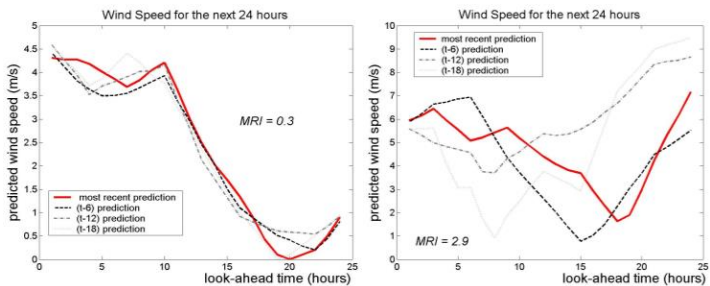

Fig. 2: *Stable (left picture) and unstable (right picture) weather situations.*

makes follows an affine empiric relation:

$$e_t^{24} = e_0 + s \cdot MRI \qquad (7)$$

which is composed by a basic part of the error $e_0$ and by a NWP-dependent error, the latest being a direct consequence of the prediction model's sensitivity to weather stability. The slope $s$ of the linear fitting model represents that sensitivity.

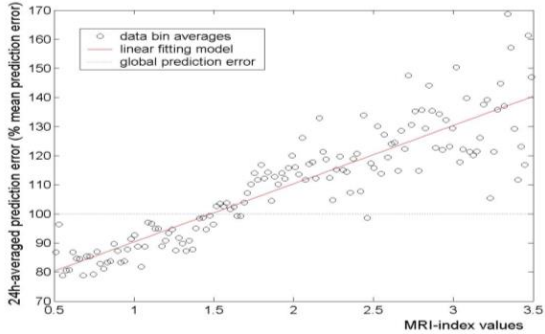

**Fig. 3:** *Prediction errors vs MRI-index over a three year dataset for a wind farm in Denmark: there is a roughly linear relation between the prediction error and MRI-index values.*

*C. On-line Adaptation of Confidence Intervals Depending on Weather Stability.*

The relation (7) indicates that when the MRI-index is low, the model is expected to be more accurate. In that case one would be ready to accept tighter confidence intervals for the predictions. The aim here is to use (7) to define a scale factor for the intervals depending on the value of the MRI-index. This scale factor can be applied to either enlarge or narrow the intervals width in the following $N_h$ hours. For instance, when the MRI-index equals 0.5, the size of the intervals for the following 24 hours is reduced by around 20%. The strategy chosen here is to only narrow the intervals when the MRI-index permits to do so. This can be done most of the times (around 65% of the times) [7].

*D. Determining the Prediction Risk from Wind Power Ensemble Forecasts*

The reasoning developed in the previous Section to quantify the meteorological risk was based on poor man's ensemble forecasts for wind speed. The idea was to quantify the meteorological risk and to determine how this relates to the error of power forecasts. In certain situations it becomes interesting to assess directly the risk from ensemble wind power forecasts. This permits to account for the effect that the power curve may have (i.e. reduced risk for wind speeds between rated and cut-off and amplified risk for wind speeds between cut-in and rated). Wind power ensemble forecasts are generated using the prediction model with input NWPs (wind speed, direction etc) provided at different time origins in the recent past. In the same way real ensemble NWPs (multiple sets of NWPs provided at the same time-origin) can be used if available.

Define $WP_{t-\gamma_i}$ to be the $N_f$ sets of wind power forecasts, with $\gamma_i$ being the age of each set. The values for $\gamma_i$ can be 0, 6, 12, etc, for the case of using Hirlam as a NWP supplier: the ensemble member $WP_{t-\gamma_i}$ is the one obtained with the meteorological forecasts of age $\gamma_i$.

Then, an index, called hereafter "production-risk" PRI-index, is defined to measure the spread of the wind power forecasts at a given time, in the same manner than the "meteo-risk" one. It uses the most recent forecast as a reference and reflects the variability of the older forecasts:

$$PRI := \sum_{i=1}^{N_f-1} p_i \cdot d\left(WP_{t-\gamma_0}, WP_{t-\gamma_i}\right) \qquad (8)$$

where the distance $d(.,.)$ and the weights $p_i$ between forecasts are similarly defined as for the MRI index in [7]. It can be easily shown that the PRI-index takes values between 0 and $P_n$. As NPRI is defined the normalized value of PRI upon $P_n$.

The meteorological risk MRI-index was seen to give information on the expected level of prediction error [7]. Similarly to this, the relation between the NPRI-index and the level of prediction error will be shown in the next Section. This relation can be used to define rules able to forecast the occurrence of outliers depending on the NPRI-index value. In an on-line environment such rules will permit to derive signals or alarms for the end-user for taking preventive actions (i.e. increase spinning reserve) or developing appropriate trading strategies.

IV. RESULTS

In this Section results are presented from the validation of the developed methodology for a wind farm in Denmark (WF-A) and for several wind farms in Ireland (WF-B to WF-F) with an installed capacity of several MWs each. The prediction model is the adaptive Fuzzy Neural Network (F-NN) model described in [3]. The available time-series cover a period of five years for WF-A, from which 12000 hours were used for training (learning set), 2000 hours for cross-validation and three years for testing the performance of the model. Regarding WF-B to WF-F, the time-series cover a period of almost two years (learning: 6600 hours, cross-validation: 1000 hours, testing: one year). The results presented here are on the testing sets.

The prediction model provides forecasts for the next 43 hours with hourly time-steps. Forecasts are updated every hour using SCADA data as input. Hirlam NWPs that are used have a spatial resolution of around 15 km for WF-A and of around 30 km for the rest. They are provided 4 times per day and at the level of wind farm as interpolated values.

*A. Confidence intervals*

Fig. 4 depicts an episode with the wind power predictions and 85% intervals for the next 36 hours compared to the real values for WF-A. This figure also illustrates the fine-tuning of the intervals: this example corresponds to a weather situation classified as stable with respect to the "meteo-risk" index. For the first 24 look-ahead times the 85% intervals are quite broad, but their size is reduced by more than 20% considering the stable expected weather situation.

Table I summarizes the observed confidence (over the 1-

year testing set) for both resampling and fine-tuned intervals. One can see that the consideration of the weather stability permits to narrow the intervals for more than 63% of the times. The average reduction is up to 11.32% of the intervals' initial size. The confidence loss due to fine-tuning is limited (col. 2).

Fig. 5 illustrates the average width of the 6-hour ahead resampling and fine-tuned intervals as a function of the confidence level. It is noted that the interval size is typical for single wind farm prediction. For the case of regional/national wind prediction some spatial smoothing effect is expected to reduce the error level.

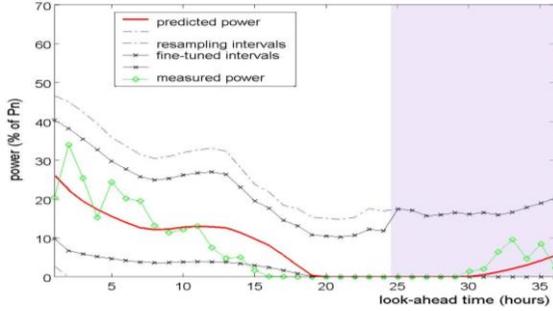

**Fig. 4:** *Wind power prediction with the resampling and fine-tuned confidence intervals for WF-F. The intervals are narrowed for the first 24 hours due to a low MRI-index value.*

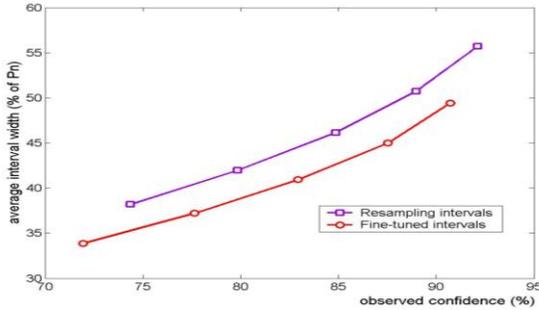

**Fig. 5:** *Average width of the 6-hour ahead confidence intervals for WF-B and for various specified confidences (75, 80, 85, 90, 95%).*

**Table I:** *Observed confidence for the two types of intervals at the end of the testing set and effects of the MRI-index on the interval reduction.*

| Wind farm | Observed confidence of resampling intervals (%) (1) | Observed confidence of fine-tuned intervals (%) (2) | No of times (%) intervals are reduced (3) | Average width reduction (%) (4) |
|---|---|---|---|---|
| WF-B | 84.87 | 82.93 | 65.10 | 11.32 |
| WF-C | 82.38 | 80.31 | 66.60 | 8.85 |
| WF-D | 81.05 | 80.53 | 68.78 | 7.28 |
| WF-E | 83.60 | 82.16 | 63.98 | 7.28 |
| WF-F | 84.29 | 82.85 | 63.52 | 8.23 |

### B. Prediction risk assessment

In order to assess the relation between the NPRI-index and the level of prediction error, we collect wind power prediction errors as obtained by the F-NN model and for the same period the NPRI-index values are estimated. The average prediction error $e_t^{24}$ for the next 24 hours, corresponding to the power forecast made at time *t*, is calculated as follows:

$$e_t^{24} = \frac{1}{24 \cdot P_n} \sum_{k=1}^{24} |e_{t+k/t}| \qquad (9)$$

These errors are then binned by NPRI-index values, and the average error $e_t^{24}$ for the next 24 hours for each bin is computed. Finally, the representative points in Fig. 6 are obtained by comparing these averages to the global prediction error $\overline{e_t^{24}}$ of the model

$$\overline{e_t^{24}} = \frac{1}{N_p} \sum_{t=1}^{N_p} e_t^{24} \qquad (10)$$

where $N_p$ is the total number of predictions made in the testing set. For each bin, the 85% confidence intervals computed by the resampling method are also given in order to visualize the errors dispersion. One can notice from this Figure that the prediction error increases with the NPRI-index, and the error dispersion too. This means that as the risk index gets higher the prediction error is likely to be greater, as well as the uncertainty on this level of prediction error.

Another way to illustrate that relation is to calculate the cumulative distribution function of the prediction errors for various bins of NPRI-index values. These curves give the probability with which an error larger than a defined threshold occurs, depending on the value of the NPRI. For instance, if at a certain time, the index takes a value between 0 and 2.5, there will be a probability of 1% that an error $e_t^{24}$ larger than the global prediction error $\overline{e_t^{24}}$ occurs. However, if at that same time the value of the index is within the [15,20) bin, the probability for the same kind of error is much larger (78%):

$$P\left(e_t^{24} > \overline{e_t^{24}} \ / \ NPRI \in [0, 2.5)\right) = 1\%$$
$$P\left(e_t^{24} > \overline{e_t^{24}} \ / \ NPRI \in [15, 20)\right) = 78\% \qquad (11)$$

Table II gives the probabilities for errors to be larger than $\frac{1}{2}$, 1, $\frac{3}{2}$ and 2 times the average error depending on the range of the NPRI-index. The Table is estimated for wind farm WF-A. Based on such a Table, several rules similar to the one given above can be derived.

Table II provides also information on the probability of extreme prediction errors to happen (extremes are defined as errors larger than twice the global prediction error of the model). Actually, for WF-A, when the NPRI-index takes low values (between 0 and 5 %) an extreme prediction error is unlikely to happen, and that is not the case if this one is within the bin [15,20) (18% probability of occurrence). On the other hand, if NPRI >10%, an error of at least 50% of the global prediction error is expected.

Finally, Fig. 7 illustrates poor man's ensemble forecasts for WF-A. In this example there are 4 ensemble members of 24 hours ahead each: the spot prediction based on the most recent NWPs and three sets of power forecasts based on Hirlam NWPs supplied 6, 12 and 18 hours ago. The NPRI-index is estimated from these ensembles and probabilities of errors larger than predefined bounds are derived. Fig. 7 illustrates two contrasting cases: the first one shows a situation where wind power predictability is quite high (low NPRI value) while the second shows a less predictable situation (high NPRI). The phase shifts between members in the second case warn on deteriorated forecasting accuracy. In fact, the average errors for the next 24 hours for the two cases are 5.83% and 15.42% respectively. The overall average error of the model for this case-study is indeed 9.23%.

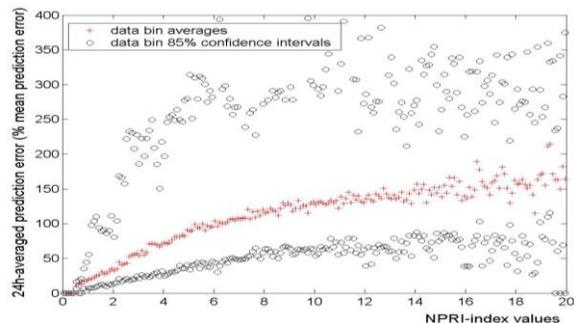

**Fig. 6:** *Prediction errors vs NPRI-index values over a 3-year dataset for a WF-A: data bin averages and 85% confidence intervals.*

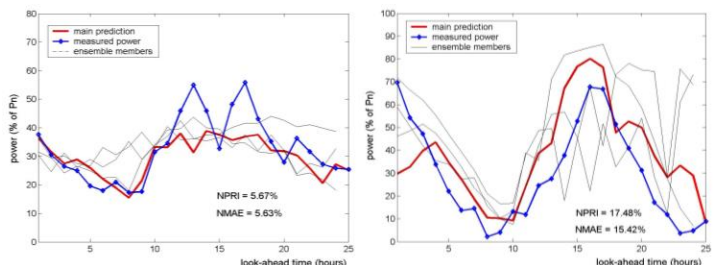

**Fig. 7:** *Wind power spot prediction and the ensemble members for WF-A (Left: NPRI= 5.67%, NMAE = 5.83% of $P_n$). (Right: NPRI= 17.48%, NMAE = 15.42% of $P_n$).*

**Table II:** *Rules for the occurrence of larger errors depending on the value of the NPRI-index for WF-A.*

| Probability (%) of occurrence of errors larger than $n$ times the global prediction error | Ranges of NPRI-index (%) | | | | |
|---|---|---|---|---|---|
| | Bin [0, 2.5) | Bin [2.5,5) | Bin [5,10) | Bin [10,15) | Bin [15,20) |
| $n = \frac{1}{2}$ | 8 | 59 | 90 | 98 | 98 |
| $n = 1$ | 1 | 13 | 42 | 65 | 78 |
| $n = \frac{3}{2}$ | 0 | 4 | 13 | 29 | 40 |
| $n = 2$ | 0 | 1 | 4 | 8 | 18 |

## V. CONCLUSIONS

A generic methodology, for assessing on-line the prediction risk of short-term wind power forecasts has been presented. It is applicable to both "physical" and "statistical" prediction models. Firstly, confidence intervals based on the resampling approach have been derived, taking into account the prediction horizon, the power class and the cut-off risk. Secondly, a new meteorological risk (MRI) index has been introduced to evaluate the weather stability. The MRI-index can be used to act on the confidence interval width afterwards depending on the weather predictability. This meteorological risk has been integrated in the wind power prediction process by producing multi-scenario wind power predictions from NWP poor man's ensemble forecasts. A second index, named as PRI, has then been derived in order to reflect the production risk. Such an index can permit to give signals to end-users on the probability of extreme prediction errors to occur.

The methodology was validated for wind farms located in Ireland and Denmark. The results are encouraging and comprise a first step in the development of on-line tools that can be used in a complementary way to the prediction model itself. The developed methods were implemented in the form of on-line modules and integrated in the Armines Wind Power Prediction System (AWPPS).

ACKNOWLEDGMENTS

The authors gratefully acknowledge ESB National Grid and ELSAM for providing data for the realization of the study.

BIOGRAPHIES

**George Kariniotakis** was born in Athens, Greece. He received his production and management engineering and M.Sc. degrees from the Technical University of Crete, Greece and his Ph.D. degree from Ecole des Mines de Paris in 1996. He is currently with the Center of Energy Studies of Ecole des Mines de Paris as an associate Scientific Manager. He is a member of IEEE. His research interests include among others renewable energies, distributed generation and computational intelligence.

**Pierre Pinson** was born in Poitiers, France. He received his applied mathematics M.Sc. degree from the National Institute for Applied Sciences (INSA Toulouse). Part of his education was made at TU Delft (The Netherlands). He is currently a Ph.D. student at the Center for Energy Studies of Ecole des Mines de Paris. His research interests include artificial intelligence, evolutionary computation and renewable energies.